\documentclass[12pt]{article}
\usepackage{amsmath}
\usepackage{cite,times}

\numberwithin{equation}{section}

\def\re{{\rm e}}
\def\rd{{\rm d}}
\def\ri{{\rm i}}
\def\re{{\rm e}}
\def\rd{{\rm d}}
\def\ri{{\rm i}}
\def\re{{\rm e}}
\def\g{{\cal G}}
\def\b{{\cal B}}
\def\pa{\partial}
\def\hp{\hphantom}

\begin{document}
\title{Lie discrete symmetries of lattice equations}
\author{Decio Levi\footnote{Permanent address: Dipartimento di Fisica,
Universit\`a Roma Tre and INFN-Sezione di Roma Tre, Via della
Vasca Navale 84, 00146-Roma, Italy}\hskip 10pt  and Miguel A.
Rodr\'{\i}guez\\ Departamento de F\'{\i}sica Te\'orica II\\
Facultad de F\'{\i}sicas, Universidad Complutense
\\ 28040-Madrid, Spain\\
e-mail: levi@fis.uniroma3.it, rodrigue@fis.ucm.es}
\maketitle
\begin{abstract}
We extend two of the methods previously introduced to find discrete
symmetries of differential equations to the case of difference and
differential-difference equations. As an example of the application of the
methods, we construct the discrete symmetries of the discrete Painlev\'e I
equation and of the Toda lattice equation.
\end{abstract}

%%%%%%%%%%%%%%%%%%%%%%%%%%%%%%%%%%%%%%%%%%%%%%%%%%%%%%%%%%%%%
%%%%%%%%%%%%%%%%%%%%%%%%%%%%%%%%%%%%%%%%%%%%%%%%%%%%%%%%%%%%%

\section{Introduction}
Symmetries have always played a very important role in the study
of differential equations \cite{Ol86}. Lie symmetries have been introduced by
Sophus Lie as a tool to unify all solution techniques for ordinary
differential equations. In particular they are useful to
\begin{itemize}
  \item get special solutions by {\sl symmetry reduction}.
  \item classify equations according to their symmetry.
  \item prove equivalence of equations under point
  transformations.
\end{itemize}

They have been extended with success to the case of
differential-difference and difference-difference equations.  As was
shown in \cite{LW93} the invariance of a differential-difference
equation with respect to a shift of the lattice $\tilde n = n + N$
provide a reduction of the equation to a system of $N$ ordinary
differential equations.  Moreover, discrete symmetries, i.e.
symmetries associated to a discrete finite group, are very important
in Quantum Mechanics.  In this field one speaks of {\sl parity, charge
conjugation, rotations by $\pi$}, etc\dots and one uses discrete
symmetries to provide selection rules.
 
Discrete symmetries are usually easy to guess but difficult to find in
a systematic way. They can be obtained by considering the normalizer of
the continuous Lie point symmetries of the equation.  This is a well
known technique and can be found in many textbooks (see for instance
\cite{GM63}).  The application to the case of differential equations
has been considered in all details by Hydon \cite{Hy98}.  The
normalizer can be obtained only in the case when a nontrivial non
commuting group of Lie point symmetries exists.

In 1996 Gaeta and Rodr\'{\i}guez \cite{GR96} introduced a modification
of Lie technique which allows us to find discrete symmetries even
when the Lie group is trivial. This technique, as we will see in
Section 3, depends on a starting ansatz and thus it does not
provide a complete result.

Reid, Weih and Wittkopf \cite{RW93} solved the determining equations
for the group transformations directly to get the discrete symmetries. 
In more recent works \cite{LR98,RB99} they simplified the work 
taking advantage of the infinitesimal results; however, the method
requires extremely heavy calculations.

In Section 2 we briefly summarize the steps necessary to obtain Lie
symmetries for difference equations while in Section 3 we present two
methods introduced to obtain discrete symmetries.  As is evident from
Section 2 all the steps necessary to obtain the discrete Lie
symmetries are not significantly different in the case of equations on
the lattice from the continuous case.  The only essential difference
is in the prolongation formula.  So in Section 4 we just present two
examples of the derivation of discrete Lie symmetries for discrete
equations.  In Section 5 one can find some concluding remarks.

%%%%%%%%%%%%%%%%%%%%%%%%%%%%%%%%%%%%%%%%%%%%%%%%%%%%%%%%%%%%%
%%%%%%%%%%%%%%%%%%%%%%%%%%%%%%%%%%%%%%%%%%%%%%%%%%%%%%%%%%%%%

\section{Point symmetries of discrete equations}
\label{s2}

Let's briefly summarize the steps necessary to obtain Lie point
symmetries for difference equations.  More details can be found in
\cite{LT01a,LT01b}.

We will consider just the case of a scalar difference equation in two
independent lattice variables as this case will cover all the examples
considered in Section 4.  The case of a differential-difference
equation is obtained from the difference-difference case by carrying
out the continuum limit in one variable.

A discrete equation of order $N_i$ in a discrete variable
$x_{n,m}^{(i)}$, $i=1,2$, is a functional relation between $N_i+1$
points in the lattice of the variable $x_{n,m}^{(i)}$. To be able to
solve the discrete equation apart from the functional relation we
need to know how the lattice points are defined. This implies that we
need four further equations in the case of two independent
variables and one for one independent lattice variable. Some of these
equations may be trivial if the lattice is orthogonal and with
constant spacing, but still are necessary to fix the symmetries and
the solutions of the difference system.

So we have:
\begin{equation}
 \label{2.1}
\Delta(x^{(k)}_{n+j,m+i},
u_{n+j,m+i}) = 0 ,
\end{equation}
\begin{equation}
 \label{2.2}
E_a(x^{(k)}_{n+j,m+i}, u_{n+j,m+i}) = 0 ,
\end{equation}
$$u_{n,m} =u(x^{(1)}_{n,m},x^{(2)}_{n,m})
,\quad 1\le a \le 4; \quad k=1, 2; \quad -i_1 \le i \le i_2; $$
$$-j_1 \le j \le j_2; \quad i_1, i_2, j_1, j_2, \in {\mathbf Z}^{\geq 0},$$
where $\Delta$ is the difference equation, $E_a$ are the equations
determining the two independent lattice variables and $n$ and $m$
are some indices which characterize them. System
(\ref{2.1},\ref{2.2}) must satisfy some obvious conditions so as
to be able to calculate the variables in all the lattice plane.

A Lie point symmetry for equations (\ref{2.1},\ref{2.2}) is a point
transformation
\begin{equation}
 \label{2.3}
\tilde x^{(i)}_{n,m} = F^{(i)}_\lambda(x^{(k)}_{n,m},  u_{n,m}),
\quad \tilde u_{n,m} = G_\lambda(x^{(k)}_{n,m}, u_{n,m}), \quad (i,k
= 1, 2 ) ,
\end{equation}
which leaves equation (\ref{2.1},\ref{2.2}) invariant. The symbol
$\lambda$ indicate the group parameters.  As this
transformation acts on the entire space where the independent and
dependent variables are defined, the functions $F^{(i)}_\lambda$
and  $G_\lambda$ determine the transformation everywhere. We can
introduce the corresponding infinitesimal transformations of
coefficients $\xi^{(i)}(x^{(k)}_{n,m},  u_{n,m})$,
$\phi(x^{(k)}_{n,m}, u_{n,m})$ and thus write the vector field
\begin{equation}
 \label{2.4}
\hat X_{n,m} = \xi^{(i)}(x^{(k)}_{n,m},  u_{n,m})
\partial_{x^{(i)}_{n,m}} + \phi(x^{(k)}_{n,m},  u_{n,m})
\partial_{u_{n,m}},
\end{equation}
and its prolongation
\begin{equation}
 \label{2.5}
\mathrm{pr}\hat X_{n,m} = \sum_{j=-j_1}^{j_2}  \sum_{l=-i_1}^{i_2} \hat
X_{n+l,m+j} .
\end{equation}

The invariance conditions, i.e., the necessary conditions which
provide the symmetries for  equations (\ref{2.1},\ref{2.2}), are given by
\begin{equation}
 \label{2.6}
\mathrm{pr}\hat X_{n,m} \Delta= 0 \qquad \text{when} \quad (\Delta, E_a)
=(0,0),
\end{equation}
\begin{equation}
 \label{2.7}
\mathrm{pr}\hat X_{n,m}E_a =0 \qquad \text{when} \quad (\Delta, E_a) =(0,0).
\end{equation}

Equations (\ref{2.6}, \ref{2.7}) are a set of equations for
$\xi^{(i)}$ and $\phi$.  In equations (\ref{2.6}, \ref{2.7})
$\xi^{(i)}$ and $\phi$ will appear in various points of the lattice,
which, when the equations $\Delta = 0$ and $E_a = 0$ have been taken
into account, are all independent.  So we have five functional
equations for the functions $\xi^{(i)}$ and $\phi$.  The variables
appearing in these five equations are all independent.  These
independent variables appear either explicitly in the equation or in
the unknown infinitesimal coefficients.  As the infinitesimal
coefficients are analytic functions we can convert the determining
equations into a system of differential equations by differentiating
them with respect to the independent variables.  In such a way we get
an overdetermined system of, in general, nonlinear partial
differential equations.  We solve the obtained equations and introduce
its solution into the functional equations and solve them.

As an example of this procedure we will
calculate the continuous symmetries of a discrete
Painlev\'e I equation \cite{LR92b}.
The discrete Painlev\'e I equation is given by
\begin{equation} \label{2.17}
u_{n+1} + u_n + u_{n-1} = \frac{\alpha x_{n} + \beta}{u_n}
+ \gamma
\end{equation}
where $\alpha$, $\beta$ and $\gamma$ are arbitrary constants and 
$u_{n}=u(x_{n})$. The
equation which defines the lattice points can be written as:
\begin{equation} \label{2.18}
x_{n+1} - x_n  = h
\end{equation}
so that $x_n = h n + x_0$. With the choice $u_n = -1
+ h^2 w(x)$, $\alpha = - h^5$, $\beta = 3$ and $\gamma = 6$
equation (\ref{2.17}) reduces, in the continuous limit, to
\begin{equation}\label{2.19}
w_{xx}=6w^2+x.
\end{equation}
which is the Painlev\'e I transcendent \cite{In56}.

The infinitesimal symmetry generator is given by
\begin{equation}
 \label{2.20}
\hat X_{n} = \xi_n(x_{n},  u_{n})\partial_{x_{n}} + \phi_n(x_{n},
u_{n})\partial_{u_{n}},
\end{equation}
and its prolongation
\begin{equation}
 \label{2.21}
\mathrm{pr}\hat X_{n,m} = \sum_{j=-1}^{1}  \hat X_{n+j} .
\end{equation}
Applying eq. (\ref{2.21}) to the lattice equation (\ref{2.18}) we
get
\begin{equation}
 \label{2.22}
\xi_{n+1}(x_{n+1},u_{n+1}) = \xi_{n}(x_{n},u_{n}),
\end{equation}
which, taking into account that $u_n$ and $u_{n+1}$ are
independent variables, gives that $\xi_n$ cannot depend from
$u_n$ and, as a function of $n$, must be a constant (cannot vary
between different points of the lattice)
\begin{equation}
 \label{2.23}
\xi_{n} = K_0.
\end{equation}
In a similar way, by applying  (\ref{2.21}) to
equation (\ref{2.17}) and differentiating the result with respect to
$u_n$ and $u_{n+1}$, one can prove that
\begin{equation}
 \label{2.24}
\phi_{n}(x_{n},u_{n}) = \phi^{(0)}_n(x_n) + u_n \phi_{n}^{(1)}(x_{n}),
\end{equation}
where $\phi^{(0)}_n$ and $\phi^{(1)}_n$ satisfy the following 
overdetermined system of equations
\begin{eqnarray}
 \label{2.25}
\phi^{(1)}_{n-1}& = &\phi_{n}^{(1)},
\\
 \label{2.26}
(\alpha x_n + \beta)(\phi^{(1)}_{n}
+ \phi_{n-1}^{(1)})&=& \alpha K_0  ,
\\
 \label{2.27}
(\alpha x_n + \beta)\phi_{n}^{(0)} &=& 0,
\\
 \label{2.28}
\phi^{(0)}_{n+1} + \phi^{(0)}_{n} + \phi^{(0)}_{n-1} + \gamma \phi^{(1)}_{n-1} &=&
0.
\end{eqnarray}

According to the values of the parameters ($\alpha$, $\beta$ and
$\gamma$) we have various possibilities:
\begin{itemize}
  \item If $\alpha \ne 0$ then from equation (\ref{2.26}) $K_0 = 0$ and
  $\phi^{(1)}_n = 0$. From equation (\ref{2.27}) $\phi^{(0)}_n = 0$ and 
  thus no symmetry is present.
  \item If $\alpha = 0$ but $\beta \ne 0$ then from
  equation (\ref{2.26}), $\phi^{(1)}_n = 0$ and from equation (\ref{2.27}), 
  $\phi^{(0)}_n =
  0$; so only $K_0 \ne 0$, i.e., only space translations are
  possible.
  \item If $\alpha = 0$, $\beta = 0$ equations (\ref{2.26}, \ref{2.27})
  are identically satisfied, equation (\ref{2.25}) implies that 
  $\phi^{(1)}_n =
  K_1$ and \cite{KP91}
  \begin{equation} \label{2.29}
  \phi^{(0)}_{n}  = - \frac{\gamma K_1}{3} + K_2 
  \left[\frac{1}{2} (\ri {\sqrt 3}-1) \right]^n +
  K_3 \left[-\frac{1}{2} (\ri {\sqrt 3}+1) \right]^n .
  \end{equation}
 So in this case the now linear equation has a four dimensional symmetry group.
\end{itemize}

To end this section let us consider the case of symmetries of
differential-difference equations.  The passage from a
difference-difference equation to a differential-difference equation
is done by carrying out the continuum limit for a lattice variable,
say $x^{(1)}_{n,m} = t_{n,m}$, when the distance between the points
along this direction goes to zero and the lattice index goes to
infinity in such a way that the position $t$ remains finite.  This
implies that the corresponding lattice spacing $\tau$ cannot be
modified by a point transformation but it is a fixed number which can
tend to zero.  So, for example, the lattice variable $t_{n,m}$ cannot
be described by dilation invariant equations like
\begin{equation}
 \label{2.8}
t_{n+1,m} - 2 t_{n,m} + t_{n-1,m} = 0, \qquad t_{n,m+1} - t_{n,m} = 0,
\end{equation}
as in this case the parameter $\tau$ is an integration constant
and not a parameter of the equations.  Let us choose,
consequently, the lattice equations for the variable $t_{n,m}$ as
\begin{equation}
 \label{2.9}
t_{n+1,m} - t_{n,m}  =  \tau , \qquad t_{n,m+1} - t_{n,m} = 0.
\end{equation}
The solution of equation (\ref{2.9}) reads
\begin{equation}
 \label{2.10}
t \equiv t_{n,m}  =  \tau n + t_0 ,
\end{equation}
where, in all generality, we can set $t_0 = 0$. As for the
remaining lattice variable $x^{(2)}_{n,m} = x_{n,m}$,   its
position is not changing in time, i.e.
\begin{equation}
 \label{2.11}
x_{n+1,m} - x_{n,m}  =  0,
\end{equation}
while its variation along the lattice, i.e., along $m$, can depend on
$t$ and on $u_{n,m}$ in a way that is defined by one of the equations
(\ref{2.2}), say $E_1=0$.

The continuous limit is obtained by considering $n \rightarrow \infty$
as $\tau \rightarrow 0$.  In such a limit the variables $x_{n,m}$ and
$u_{n,m}$ will no more depend on $n$, equation (\ref{2.1}), if it has
the proper dependence from $\tau$, will reduce to a
differential-difference equation for $u_m = u(t,x_m)$ and equations
(\ref{2.2}) will reduce to just one equation for the lattice variable
$x_m$, $E_1 = 0$, while the other equations (\ref{2.9}, \ref{2.11})
are identically satisfied in the limit.

In this limit the symmetry vector $\hat X_{n,m}$, given by equation
(\ref{2.5}), reduces to $\hat X_m$, which will inherit the properties
of $\hat X_{n,m}$ as applied to the lattice equations (\ref{2.9},
\ref{2.11}) even if these equations in the limit reduce to $0 \equiv
0$.  Applying equation (\ref{2.6}) to equations (\ref{2.9},
\ref{2.11}), with $x^{(1)}_{n,m} = t_{n,m}$ and $x^{(2)}_{n,m} =
x_{n,m}$, we get the following three determining equations
\begin{eqnarray}
 \label{2.12}
\xi^{(1)}(t_{n,m} + \tau, x_{n+1,m}, u_{n+1,m}) &= &\xi^{(1)}(t_{n,m},
x_{n,m}, u_{n,m}) ,
\\
 \label{2.13}
\xi^{(1)}(t_{n,m}, x_{n,m+1}, u_{n,m+1}) &=& \xi^{(1)}(t_{n,m},
x_{n,m}, u_{n,m}) ,
\\
 \label{2.14}
\xi^{(2)}(t_{n,m} + \tau, x_{n+1,m}, u_{n+1,m}) &=& \xi^{(2)}(t_{n,m},
x_{n,m}, u_{n,m}) .
\end{eqnarray}
As the differential-difference equation will involve at least
$u_{n+1,m}$, $u_{n,m}$, $u_{n,m+1}$, we can always take $u_{n,m},
u_{n,m+1}$ as independent variables and express $u_{n+1,m}$ in
term of them. By differentiating equation (\ref{2.12}) with
respect to $u_{n,m+1}$ we get $\xi^{(1)}(t_{n,m}, x_{n,m})$. By a
similar reasoning for the variable $x_{n,m}$  we can reduce the
function $\xi^{(1)}$ to $\xi^{(1)}(t_{n,m})$, i.e., the function
$\xi^{(1)}$ is just a function of $t$. In a similar way we will
find that
\begin{equation}
 \label{2.15}
\hat X_{n,m} = \xi^{(1)}(t_{n,m}) \partial_{t_{n,m}} +
\xi^{(2)}(t_{n,m}, x_{n,m}) \partial_{x_{n,m}} + \phi(t_{n,m},
x_{n,m},  u_{n,m}) \partial_{u_{n,m}},
\end{equation}
and thus, in the continuum limit, we must have
\begin{equation}
 \label{2.16}
\hat X_{m} = \xi^{(1)}(t) \partial_{t} + \xi^{(2)}(t, x_{m})
\partial_{x_{m}} + \phi(t, x_{m},  u_{m}) \partial_{u_{m}}.
\end{equation}

%%%%%%%%%%%%%%%%%%%%%%%%%%%%%%%%%%%%%%%%%%%%%%%%%%%%%%%%%%%%%
%%%%%%%%%%%%%%%%%%%%%%%%%%%%%%%%%%%%%%%%%%%%%%%%%%%%%%%%%%%%%

\section{The calculus of discrete symmetries}

We will now discuss two methods of determining the discrete symmetries
of differential equations.  One of them was proposed by Hydon
\cite{Hy98,Hy00a} and it is essentially the classical method of
constructing the normalizer of a group.  The other, due to Gaeta and
Rodr\'{\i}guez \cite{GR96}, is a modification of Lie's method,
defining a discrete symmetry as a discretization of the parameter of a
continuous symmetry.  There are at least two other methods to construct
discrete symmetries.  One of them is based in a formulation of
differential equations through differential forms \cite{KC01} and the
other \cite{RW93,LR98,RB99} solves the determining equations for the
group transformations directly.  They will not be considered in this
work.

%%%%%%%%%%%%%%%%%%%%%%%%%%%%%%%%%%%%%%%%%%%%%%%%%%%%%%%%%%%%%

\subsection{Automorphisms of the symmetry algebra}

Let us consider a differential equation for a dependent variable $u$
and $N$ independent variables ${\mathbf x}=(x_{1},\ldots,x_{N})$, with
a Lie group of symmetries, $G$ and its corresponding Lie algebra $\g$. 
We construct all the automorphisms of this Lie algebra.  Some of them
will, obviously, correspond to continuous symmetries.  Others will be
essentially new and will define discrete symmetries of the equation. 
And, finally, others will not be symmetries of the equation.  These
ideas are very well known in the theory of Lie algebras and groups. 
Inner automorphisms correspond to conjugation by elements of the
group, while outer automorphisms have not this character.  See for
instance \cite{GM63} where the complete and general Lorentz group are
obtained in this way from the proper Lorentz group.

Let $\g$ be a Lie algebra of finite dimension $n$ and $\b=\{X_{1},
\ldots, X_{n}\}$ a basis of $\g$. The commutation relations in
this basis can be written as:
\begin{equation}
[X_{i},X_{j}]=c_{ij}^k X_{k},
\end{equation}
where $c_{ij}^k$ are the structure constants of $\g$ (a sum is
understood over repeated indices running from 1 to $n$).

The defining equation for an automorphism $\phi:\g\longrightarrow
\g$ is
\begin{equation}
\phi[X,Y]=[\phi(X),\phi(Y)],
\end{equation}
which,  in the basis $\b$, reads:
\begin{equation}
\phi[X_{i},X_{j}]=c_{ij}^k\phi(X_{k})=
[\phi(X_{i}),\phi(Y_{j})].\label{auto}
\end{equation}

Let $\Phi$ be the matrix representation of $\phi$ in the basis 
$\b$ ($\det\Phi\neq
0$). Then
\begin{equation}
\phi(X_{i})=\Phi^j_{\hp{j}i}X_{j}
\end{equation}
and eq. (\ref{auto}) is written as:
\begin{equation} \label{3.5}
c_{ij}^k \Phi^l_{\hp{l}k}=
\Phi^m_{\hp{m}i}\Phi^r_{\hp{j}j}c_{mr}^l
\end{equation}
for all indices $i,j,l$ (the equation is skewsymmetric in $i,j$).
If we define the matrices of the adjoint representation
\begin{equation}
C(i)^{k}_{\hp{k}j}=c_{ij}^k,
\end{equation}
 equation (\ref{3.5}) can also be written as:
\begin{equation}
\Phi C(i)=\Phi^l_{\hp{l}i}C(l)\Phi\label{ec}.
\end{equation}
Automorphisms which are related through conjugation by an element
of the symmetry group will be considered equivalent. Thus the
matrix $\Phi$ can be simplified by conjugations with the symmetry
transformations (at least if the algebra is not Abelian). This
can be done in terms of the adjoint representation.

Let us consider those elements $g_i\in G$ which are generated by an
element $X_{i}$ of the basis $\b$ of the corresponding Lie algebra
$\g$,
\begin{equation}
    g_i=\re^{\lambda X_{i}}.
\end{equation}
The conjugation provides the transformation
\begin{equation}
    \phi\rightarrow \re^{-\lambda X_{i}}\phi\,\re^{\lambda X_{i}}.
\end{equation}
We can consider the transformation of the elements of the basis
$\b$ and we have:
\begin{equation}
    \re^{-\lambda X_{i}}X_{j}\,\re^{\lambda
    X_{i}}=A(i,\lambda)^k_{\hp{k}j}X_{k}.
    \end{equation}
It is not difficult to show that
\begin{equation}\label{transf}
A(i,\lambda)=\re^{\lambda C(i)}.
\end{equation}
We can apply the transformations (\ref{transf}):
\begin{equation}
 \Phi\rightarrow  \re^{\lambda C(i)} \Phi
\end{equation}
to simplify the matrix $\Phi$. 

We can now construct a representation of the automorphism in the space
of variables.  Let $X$ be a vector field given by (the indices $a,b$
run from 1 to $N$)
\begin{equation}\label{vectfd}
 X=\xi^{a}({\mathbf x},u)\pa_{x_{a}}+\varphi({\mathbf x},u)\pa_{u}
\end{equation}
which is a symmetry of the equation under study, and let us consider a
symmetry, given by the transformation:
\begin{equation}
   ({\mathbf x},u)\rightarrow (\hat{\mathbf x}(x,u),\hat{u}(x,u))
\end{equation}
   The vector field $X$ is transformed into a new vector field
\begin{equation}
 \hat{X}=\hat{\xi}^{a}(\hat{\mathbf x},\hat{u})\pa_{\hat{x}_{a}}+
 \hat{\varphi}(\hat{\mathbf x},\hat{u})\pa_{\hat{u}}.
\end{equation}
As the transformation is a symmetry of the equation, $\hat X$
must have the same form in the new variables:
\begin{equation}
 \hat{X}=\xi^{a}(\hat{\mathbf x},\hat{u})\pa_{\hat{x}_{a}}+
 \varphi(\hat{\mathbf x},\hat{u})\pa_{\hat{u}}.
\end{equation}

If we consider a basis of the symmetry algebra, $\{X_{i}\},\, 
i=1,\ldots, n$, the
transformed vector fields are:
\begin{equation}
 \hat{X}_{i}=\xi^{a}_{i}(\hat{\mathbf x},\hat{u})\pa_{\hat{x}_{a}}+
 \varphi_{i}(\hat{\mathbf x},\hat{u})\pa_{\hat{u}}
\end{equation}
and, as the transformation is an automorphism of the algebra,
\begin{equation}
   \hat{X}_{i}=\Phi^j_{\hp{j}i}X_{j}.\label{trans}
    \end{equation}
Applying (\ref{trans}) to the new variables, we get:
\begin{equation}
   \hat{X}_{i} \hat{x}_{a}=\xi^{a}_{i}(\hat{\mathbf x},\hat{u})=
   \Phi^j_{\hp{j}i}X_{j}\hat{x}_{a}
   =\Phi^j_{\hp{j}i}\left(\xi^{b}_{j}({\mathbf x},u)
   \frac{\pa \hat{x}_{a}}{\pa x_{b}}+\varphi_{j}({\mathbf
   x},u)\frac{\pa\hat{x}_{a}}{\pa u} \right)
    \end{equation}
and
\begin{equation}
   \hat{X}_{i} \hat{u}=\Phi_{i}(\hat{\mathbf x},\hat{u})=
   \Phi^j_{\hp{j}i}X_{j} \hat{u}
   =\Phi^j_{\hp{j}i}\left(\xi^{b}_{j}({\mathbf x},u)
   \frac{\pa \hat{u}}{\pa x_{b}}+\varphi_{j}({\mathbf
   x},u)\frac{\pa \hat{u}}{\pa u} \right)
    \end{equation}
or:
\begin{eqnarray}\label{fineq}
\xi^{b}_{j}({\mathbf x},u)
   \frac{\pa \hat{x}_{a}}{\pa x_{b}}+\varphi_{j}({\mathbf
   x},u)\frac{\pa\hat{x}_{a}}{\pa u}&=&
   (\Phi^{-1})^i_{\hp{i}j} \xi^{a}_{i}(\hat{\mathbf x},\hat{u})\\
\xi^{b}_{j}({\mathbf x},u)
   \frac{\pa \hat{u}}{\pa x_{b}}+\varphi_{j}({\mathbf
   x},u)\frac{\pa \hat{u}}{\pa u}&=&
   (\Phi^{-1})^i_{\hp{i}j}\varphi_{i}(\hat{\mathbf 
   x},\hat{u}).\nonumber
\end{eqnarray}
We have to solve eqs. (\ref{fineq}) to find the expression of
the automorphism as a transformation in our space of variables and
functions. After, we must check if the automorphism is a
symmetry of the equation and it does not correspond to a
continuous symmetry.

%%%%%%%%%%%%%%%%%%%%%%%%%%%%%%%%%%%%%%%%%%%%%%%%%%%%%%%%%%%%%

\subsection{Determining equation for a discrete symmetry}

We will now consider the method of the 
discretization of the parameter of a continuous transformation. Let us
briefly review how the method works. To simplify the description
(see \cite{GR96} for a detailed exposition), let us consider a
smooth curve in ${\mathbf R}^2$, $y=f(x)$ and a vector field
\begin{equation}
X=\xi(x,y)\partial_x+\varphi(x,y)\partial_y
\end{equation}
The point $(x,y)$ is transformed under the infinitesimal action as
\begin{equation}
x'=x+\lambda \xi(x,y),\quad y'=y+\lambda\varphi(x,y)
\end{equation}
and the graph of our curve is transformed into a new one, defined
by the transformed function:
\begin{equation}
f_{\lambda}(x)=f(x)+\lambda[\varphi(x,y)-\xi(x,y)\partial_xf(x)]\label{inft}
\end{equation}
If we introduce a function $F(x;\lambda)$ such that $F(x;\lambda)=
f_{\lambda}(x)$, $F$, taking into account (\ref{inft}) will satisfy 
the partial differential equation
\begin{equation}
\frac{\partial F(x;\lambda)}{\partial\lambda}+\xi(x,F(x;\lambda))
\frac{\partial F(x;\lambda)}{\partial
x}=\varphi(x,F(x;\lambda)).\label{detmg}
\end{equation}
In order to recover the original function, we have to impose the
initial condition:
\begin{equation}\label{3.27}
F(x;0)=f(x)
\end{equation}
and if we want to obtain the same function for a particular value
$\lambda_0$ of the parameter and hence a discrete symmetry
\begin{equation}\label{per}
F(x;\lambda_{0})=F(x;0)=f(x)
\end{equation}
This is equivalent to find periodic solutions of the equation
(\ref{detmg}). This equation is a functional equation and hard to
solve.

If we consider a differential equation instead of the graph of a
function, we have to pose the same question in the appropriated
jet space \cite{Ol86}. The answer is simpler in this case, as the
right hand side of equation (\ref{detmg}) is computed from the
prolongation of the vector field under consideration.

Let us consider a differential equation:

\begin{equation}\label{3.29}
\partial_{J}u=f({\mathbf x},u,\partial_{J'}u,\ldots)
\end{equation}
where
$\partial_{J}u=\partial_{x_{1}}^{j_{1}}\ldots
\partial_{x_{N}}^{j_{N}}u$,
$J=(j_{1},\ldots,j_{N})$.

The determining equation, used to describe the discrete
symmetries, is
then:
\begin{equation}
\frac{\partial F}{\partial \lambda}+\sum_i
\xi_{i}\frac{\partial F}{\partial x_{i}}+
\sum_{J'}\phi^{J'}\frac{\partial F}{\partial u_{J'}}=
\phi^{J}\bigg|_{\partial_{J}u=F}
\end{equation}
where $\phi^{J'}$ are the $J'$-prolongations of the vector field
(\ref{vectfd}) and one has to rewrite equation (\ref{3.29})
in terms of $F$. Then one looks for periodic solutions of this 
equation in $\lambda$.

For instance, consider a differential equation of the
following type:
\begin{equation}
u_{tt}=f({\mathbf x},u,u_{t}).
\end{equation}
Choosing $F=F({\mathbf x},u,u_{t};\lambda)$, the determining equation is:
\begin{equation}\label{deter}
\frac{\partial F}{\partial \lambda}+\sum_i
\xi_{i}\frac{\partial F}{\partial x_{i}}+
\phi\frac{\partial F}{\partial u}+
\phi^{t}\frac{\partial F}{\partial u_{t}}=
\phi^{tt}\bigg|_{\partial_{tt}u=F}
\end{equation}
and
\begin{equation}
F({\mathbf x},u,u_{t};0)=F({\mathbf x},u,u_{t};\lambda_0)=
f({\mathbf x},u,u_{t})
\end{equation}
for some $\lambda_0$.

To apply this method we do not need to know the continuous symmetries. 
In fact, the determining equation provides a general solution which
does not depend on the differential equation we are studying.  The
equation appears when we impose the boundary condition as in
(\ref{3.27}) and the discrete symmetries by requiring equation
(\ref{per}).  However, we have to make some assumptions to get a
solution of the equation, because we do not know the symmetries
which appear in it.  We will use this method to determine the discrete
symmetries of the Painlev\'e I equation (\ref{2.17}).

%%%%%%%%%%%%%%%%%%%%%%%%%%%%%%%%%%%%%%%%%%%%%%%%%%%%%%%%%%%%%
%%%%%%%%%%%%%%%%%%%%%%%%%%%%%%%%%%%%%%%%%%%%%%%%%%%%%%%%%%%%%

\section{Lie discrete symmetries of lattice equations}

The construction shown in Section 2 of point symmetries of discrete
equations is very similar to the standard approach to point
symmetries of continuous equations \cite{Ol86}.  The main difference
lies in the form of the prolongation and in the way we solve the
determining equations.  Consequently, the procedure presented in
Section 3 for constructing discrete symmetries of continuous equations
can be carried over in a straightforward manner to the discrete case,
just by changing the form of the prolongation.  So, in the following
we will just apply the methods discussed in Section 3 to two examples
of equations of Mathematical Physics, the discrete Painlev\'e equation
(\ref{2.17}) and the Toda equation \cite{To81}.  The Painlev\'e
equation, as we have shown in Section 2, has no continuous symmetry
(for a generic choice of the parameters), and we will use the
determining equation method to find its discrete symmetries.  In the
Toda equation we have a continuous group of symmetries
\cite{LW91,HL00} and we will construct the discrete symmetries by
computing the automorphisms of the corresponding Lie algebra.

%%%%%%%%%%%%%%%%%%%%%%%%%%%%%%%%%%%%%%%%%%%%%%%%%%%%%%%%%%%%%

\subsection{Discrete Painlev\'e equation}

As we have said in the introduction to this section, the discrete
Painlev\'e I equation (\ref{2.17}) has no continuous symmetry so that
we cannot apply the method of the normalizer.
 
We define the extended $\lambda$ dependent equation in the
transformed space as
\begin{equation} \label{4.1}
{\hat u}_{+} + {\hat u}_{-} = F({\hat x}, {\hat u};
\lambda),
\end{equation}
where, for $\lambda = 0$
\begin{equation} \label{4.1b}
F({\hat x}, {\hat u}; 0) = -  u + \frac{\alpha x
+ \beta}{u} + \gamma.
\end{equation}
For the sake of simplicity we have used the notation
\begin{eqnarray*}
&&x=x_{n},\quad x_{+}=x_{n+1},\quad 
x_{-}=x_{n-1},\\
&&u=u(x_{n}),\quad u_{+}=u(x_{n+1}),\quad 
u_{-}=u(x_{n-1}).
\end{eqnarray*}

The generic Lie point transformation, written in terms of the infinitesimal
symmetry generators, is:
\begin{eqnarray} \label{4.2}
\frac {d {\hat x}}{d \lambda} = \xi({\hat x},{\hat u}),
\qquad  \frac {d {\hat u}}{d \lambda} = \phi({\hat x},{\hat
u}).
\end{eqnarray}
We consider the same definition of the lattice (\ref{2.18}) that we
used for constructing the Lie point symmetries in Section 2.  So we
have $\xi=K_{0}$.  Differentiating equation (\ref{4.1}) with respect
to $\lambda$ and taking into account equation (\ref{4.2}) we get
(\ref{deter}):
\begin{eqnarray}
\frac{\partial F}{\partial \lambda}+ K_{0}\frac{\partial
F}{\partial {\hat x}}+ \phi({\hat x},{\hat u})\frac{\partial
F}{\partial {\hat u}}= \phi({\hat x}_{+},{\hat u}_{+})  +
\phi({\hat x}_{-},  F- {\hat u}_{+} ) \label{4.3}
\end{eqnarray}
By differentiating equation (\ref{4.3}) with respect to ${\hat u}_{+}$
and ${\hat u}$ we get
\begin{equation} \label{4.3b}
\phi({\hat x}, {\hat u})= \phi^{(0)}({\hat x}) + {\hat u}\,
\phi^{(1)}({\hat x}).
\end{equation}
Substituting equation (\ref{4.3b}) into equation (\ref{4.3}) and requiring
that the obtained equation be satisfied for any $u_{+}$, we get
\begin{equation} \label{4.3bc}
\phi^{(1)}(\hat{x}) = K_1 + K_2 (-1)^{\frac{{\hat x}-\hat{x}_{0}}{h}}.
\end{equation}
Equation (\ref{4.3}) is now reduced to a linear partial
differential equation of first order which can be solved on the
characteristics:
\begin{eqnarray}\label{4.4}
\frac{\rd\lambda}{1}= \frac{\rd {\hat x}}{K_{0}}= \frac{\rd {\hat
u}}{\phi^{(0)} + {\hat u}\, \phi^{(1)}}= \frac{\rd F}{\phi^{(0)}_{+}
  + \phi^{(0)}_{-} + \phi^{(1)}_{-}
F }\, ,
\end{eqnarray}
where
\begin{equation}
\phi^{(i)}=\phi^{(i)} (x),\quad \phi^{(i)}_{+}=\phi^{(i)} (x_{+}),\quad 
\phi^{(i)}_{-}=\phi^{(i)} (x_{-}),\quad i=0,1.
\end{equation}
The first invariant is:
\begin{equation}\label{4.5}
{\hat x}=x + K_{0} \lambda.
\end{equation}
The action of this transformation on the discrete Painlev\'e
equation (\ref{2.17}) corresponds just to a change in the
parameter $\beta$. We could carry out the calculation with 
$K_{0}\neq 0$, but for the sake of clarity of the presentation, we 
have set $K_{0}=0$ as this transformation cannot provide any
discrete Lie transformation. Consequently, $\phi^{(i)}$ will not 
depend on $\lambda$.

The next invariant is obtained by integrating equation (\ref{4.4}) for
${\hat u}$ as a function of $\lambda$. We have:
\begin{equation}\label{4.6}
{\hat u}=\re^{\lambda\phi^{(1)}  }\left[ u  + \frac
{\phi^{(0)}} {\phi^{(1)}}\left( 1 - \re^{ -
 \lambda\phi^{(1)} } \right)\right] 
\end{equation}
The integration of equation (\ref{4.4}) for $F$ satisfying the boundary
conditions (\ref{4.1b}) gives
\begin{eqnarray}\label{4.7}
F({\hat x}, {\hat u};\lambda)& =& \re^{\lambda \phi^{(1)}_{-}
}\bigg[  \frac {\phi^{(0)}}
{\phi^{(1)}} ( 1 - \re^{-\lambda\phi^{(1)}  }) +
\frac{\alpha {\hat x} + \beta}{{\hat u} \re^{-
\phi^{(1)} \lambda } - \frac {\phi^{(0)}}
{\phi^{(1)}} ( 1
- \re^{-\lambda\phi^{(1)}  })} + \gamma  \bigg] \nonumber\\ &&
- {\hat u}\re^{\lambda [\phi^{(1)}_{-}
-\phi^{(1)}]  }  - \frac {\phi^{(0)}_{+}+ \phi^{(0)}_{-}}
{\phi^{(1)}_{-}}(1-\re^{\lambda \phi^{(1)}_{-}}).
\end{eqnarray}

To get a discrete Lie symmetry we require that there exists a
value of $\lambda$, say $\lambda_0$, such that
\begin{equation} \label{4.8}
F({\hat x}, {\hat u};0) = F({\hat x}, {\hat u};
\lambda_0).
\end{equation}
If we want (\ref{4.8}) to be satisfied, we need
\begin{eqnarray}
\label{one}   && (\phi^{(1)}_{-}-\phi^{(1)})\lambda_{0}=2\pi\ri N,\quad 
N\in {\mathbf Z}\\
\label{two}&& \phi^{(0)}(1-\re^{-\lambda_{0}\phi^{(1)}})=0
\end{eqnarray}

From (\ref{one}) we obtain $K_{2}=0$ and then, $\phi^{(1)}=K_{1}$. 
Equation (\ref{two}) is solved by requiring one or the other of the following 
two conditions:
\begin{eqnarray}
\label{three}    &&\lambda_{0}K_{1}=2\pi\ri N,\quad N\in {\mathbf Z}\\
\label{four} && \phi^{(0)}=0
\end{eqnarray}
In the case of condition (\ref{three}), equation (\ref{4.8}) is 
satisfied. However, this provides no discrete symmetry.

Let us go over to the second condition (\ref{four}).
If $\gamma\neq 0$, equation (\ref{4.8}) implies equation 
(\ref{three}), i.e.,  no discrete symmetry is present.
If $\gamma = 0$, then we have $K_{1}\lambda_0 = \ri \pi N$,
providing a discrete symmetry ${\hat u} = \pm u$ even in the case when
no continuous symmetry is present.  However, in this case, the
continuum limit of this difference equation is not Painlev\'e I.

%%%%%%%%%%%%%%%%%%%%%%%%%%%%%%%%%%%%%%%%%%%%%%%%%%%%%%%%%%%%%

\subsection{Discrete symmetries of the Toda equation}
Let us consider as our second example the Toda equation:
\begin{equation}
u_{tt}=\re^{u_{+}-u}-\re^{u-u_{-}}\label{toda}
\end{equation}
where $x_{\pm}=x\pm h$, $u_{\pm}=u(x\pm h,t)$ and
$h$ is the lattice step (see equation (\ref{2.18}).

As it is well known \cite{LW91,LR92a,HL00}, the following operators form a
basis of the symmetry algebra of the Toda equation:
\begin{equation}
X_{1}=\pa_{u},\quad X_{2}=\pa_{x},\quad
X_{3}=\pa_{t},\quad X_{4}=t\pa_{u},\quad X_{5}=
t\pa_{t}-\frac{2x}{h}\pa_{u}\label{base}\end{equation}
The nonzero commutation relations are:
\begin{equation}
[X_{2},X_{5}]=-\frac{2}{h}X_{1},\quad [X_{3},X_{4}]=X_{1},\quad
[X_{3},X_{5}]=X_{3},\quad [X_{4},X_{5}]=-X_{4}
\end{equation}
and the matrices $C(i)$ of the adjoint representation are:
\begin{eqnarray}
C(1)=0,\;
C(2)=\left(\begin{array}{ccccc}
\cdot &\cdot&\cdot&\cdot& -2/h\\
\cdot &\cdot&\cdot&\cdot&\cdot \\
\cdot &\cdot&\cdot&\cdot&\cdot \\
\cdot &\cdot&\cdot&\cdot&\cdot \\
\cdot &\cdot&\cdot&\cdot&\cdot \end{array}\right), \;
C(3)=\left(\begin{array}{ccccc}
\cdot &\cdot&\cdot&1& \cdot\\
\cdot &\cdot&\cdot&\cdot&\cdot \\
\cdot &\cdot&\cdot&\cdot&1 \\
\cdot &\cdot&\cdot&\cdot&\cdot \\
\cdot &\cdot&\cdot&\cdot&\cdot \end{array}\right),\nonumber\\
C(4)=\left(\begin{array}{ccccc}
\cdot &\cdot&-1&\cdot&\cdot\\
\cdot &\cdot&\cdot&\cdot&\cdot \\
\cdot &\cdot&\cdot&\cdot&\cdot \\
\cdot &\cdot&\cdot&\cdot&-1 \\
\cdot &\cdot&\cdot&\cdot&\cdot \end{array}\right),\;
C(5)=\left(\begin{array}{ccccc}
\cdot &2/h &\cdot&\cdot&\cdot\\
\cdot &\cdot&\cdot&\cdot&\cdot \\
\cdot &\cdot&-1&\cdot&\cdot\\
\cdot &\cdot&\cdot&1&\cdot \\
\cdot &\cdot&\cdot&\cdot&\cdot \end{array}\right).
\end{eqnarray}

Applying equation (\ref{ec}) we get :
\begin{equation}
b_{21}=b_{23}=b_{24}=b_{31}=b_{41}=b_{51}=b_{53}=b_{54}=0,\label{cero}
\end{equation}
where, to simplify the notation we have written 
$\Phi^i_{\hp{i}j}=b_{ij}$.
The most significant remaining equations are:
\begin{eqnarray}
    && b_{34}(b_{55}+1)=b_{44}(b_{55}-1)=
    b_{33}(b_{55}-1)=b_{43}(b_{55}+1)=0,\label{uno}\\
    && b_{11}=b_{33}b_{44}-b_{34}b_{43}.\label{dos}
    \end{eqnarray}
The determinant of $\Phi$ must be different from zero ($\phi$ is an
automorphism) and consequently, taking into account equation 
(\ref{cero})), we must have $b_{11}\neq
0$. Using equation (\ref{uno}) we conclude that
$b_{55}=\pm 1$. We will distinguish two cases
\begin{description}
\item{a)}  $b_{55}=1$
In this case, $b_{34}=b_{43}=0$ and $b_{33}\neq 0$, $b_{44}\neq
0$. The other elements of the matrix $\Phi$ must satisfy the equations:
\begin{eqnarray}
&&b_{32}=b_{42}=b_{52}=0,\nonumber\\\
&&b_{13}=b_{33}b_{45},\quad b_{14}=b_{35}b_{44},\\
&&b_{11}=b_{22}=b_{33}b_{44}.\nonumber
\end{eqnarray}

The matrix $\Phi_{1}=\Phi(b_{55}=1)$ is:
\begin{equation}
    \Phi_{1}=\left(\begin{array}{ccccc}
    b_{33}b_{44} & b_{12} & b_{33}b_{45} & b_{35}b_{44} & b_{15}\\
    0   & b_{33}b_{44} & 0 & 0 &  b_{25}\\
    0 & 0 & b_{33} & 0 & b_{35}\\
   0 & 0 & 0 & b_{44} & b_{45} \\
   0 & 0 & 0 & 0 & 1  \end{array}\right).
\end{equation}

\item{ b)}  $b_{55}=-1$
Now, $b_{33}=b_{44}=0$ and $b_{34}\neq 0$, $b_{43}\neq 0$. The
other elements satisfy the equations:
\begin{eqnarray}
&&b_{32}=b_{42}=b_{52}=0,\nonumber\\
&&b_{13}=-b_{35}b_{43},\quad b_{14}=-b_{34}b_{45},\\
&&b_{11}=-b_{22}=-b_{34}b_{43}.\nonumber
\end{eqnarray}

The matrix $\Phi_{2}=\Phi(b_{55}=-1)$ is:
\begin{equation}
    \Phi_{2}=\left(\begin{array}{ccccc}
    -b_{34}b_{43} & b_{12} & -b_{35}b_{43} & -b_{34}b_{45} & b_{15}\\
    0   & b_{34}b_{43} & 0 & 0 &  b_{25}\\
    0 & 0 & 0 & b_{34} & b_{35}\\
   0 & 0 & b_{43} & 0 & b_{45} \\
   0 & 0 & 0 & 0 & -1  \end{array}\right).
\end{equation}
\end{description}
To simplify the matrix $\Phi$ we will conjugate the automorphism
using the continuous transformations in the adjoint
representation.

The exponentials of the matrices $C(i)$ are easy to find. Using $C(3)$ 
we can put $b_{35}=0$, with $C(4)$, $b_{45}=0$, 
$C(5)$ gives $b_{12}=0$, and, finally using $C(2)$, $b_{15}=0$. Then, the
simplified $\Phi_{1}$ is:
\begin{equation}
    \Phi_{1}=\left(\begin{array}{ccccc}
    b_{33} b_{44} & 0 & 0 & 0 & 0\\
    0  & b_{33}b_{44} & 0 & 0 &  b_{25}\\
    0 & 0 & b_{33} & 0 & 0\\
    0 & 0 & 0 & b_{44} & 0 \\
    0 & 0 & 0 & 0 & 1  \end{array}\right),\quad b_{33},\,b_{44}\neq 0
\end{equation}

The same procedure can be used with $\Phi_{2}$ and the result is:
\begin{eqnarray} \Phi_{2}=\left(\begin{array}{ccccc} -b_{34}b_{43}
&0&0&0&0\\ 0&b_{34}b_{43}&0&0&b_{25}\\ 0&0&0&b_{34}&0\\
0&0&b_{43}&0&0\\ 0&0&0&0&-1 \end{array}\right),\quad
b_{34},\,b_{43}\neq 0.
\end{eqnarray}

The following step  is to realize the
automorphisms in the space of the variables and functions of the 
Toda equation. We have to solve the following
system of equations:
\begin{eqnarray}
\tau_{j}(t,x,u)
   \frac{\pa \hat{t}}{\pa t}+
   \xi_{j}(t,x,u)
   \frac{\pa \hat{t}}{\pa x}+
   \varphi_{j}(t,x,u)
   \frac{\pa\hat{t}}{\pa u}=
   (\Phi^{-1})^i_{\hp{i}j} \tau_{i}(\hat{t},\hat{x},\hat{u}),\nonumber\\
\tau_{j}(t,x,u)
   \frac{\pa \hat{x}}{\pa t}+
   \xi_{j}(t,x,u)
   \frac{\pa \hat{x}}{\pa x}+
   \varphi_{j}(t,x,u)
   \frac{\pa\hat{x}}{\pa u}=
   (\Phi^{-1})^i_{\hp{i}j} \xi_{i}(\hat{t},\hat{x},\hat{u}),\\
\tau_{j}(t,x,u)
   \frac{\pa \hat{u}}{\pa t}+
   \xi_{j}(t,x,u)
   \frac{\pa \hat{u}}{\pa x}+
   \varphi_{j}(t,x,u)
   \frac{\pa\hat{u}}{\pa u}=
   (\Phi^{-1})^i_{\hp{i}j} \varphi_{i}(\hat{t},\hat{x},\hat{u}),\nonumber
    \end{eqnarray}
where  $j=1,\ldots, 5$. In Case a),
\begin{equation}
    \Phi_{1}^{-1}=\left(\begin{array}{ccccc}
    \mu\nu & 0 & 0 & 0 & 0\\
    0   & \mu\nu & 0 & 0 &  \sigma\\
    0 & 0 & \mu & 0 & 0\\
   0 & 0 & 0 & \nu & 0 \\
   0 & 0 & 0 & 0 & 1  \end{array}\right),\quad \mu,\, \nu\neq 0.
\end{equation}
Then, for $j=1$, $X_{1}=\pa_{u}$
\begin{equation}
\frac{\pa\hat{t}}{\pa u}=0,\quad
\frac{\pa\hat{x}}{\pa u}=0,\quad
\frac{\pa\hat{u}}{\pa u}=\mu\nu.
\label{equ}\end{equation}
For $j=2$, $X_{2}=\pa_{x}$
\begin{equation}
\frac{\pa\hat{t}}{\pa x}=0,\quad
\frac{\pa\hat{x}}{\pa x}=\mu\nu,\quad
\frac{\pa\hat{u}}{\pa x}=0.
\label{eqd}\end{equation}
For $j=3$, $X_{3}=\pa_{t}$
\begin{equation}
\frac{\pa\hat{t}}{\pa t}=\mu,\quad
\frac{\pa\hat{x}}{\pa t}=0,\quad
\frac{\pa\hat{u}}{\pa t}=0.
\label{eqt}\end{equation}
For $j=4$, $X_{4}=t\pa_{u}$
\begin{equation}
t\frac{\pa\hat{t}}{\pa u}=0,\quad
t\frac{\pa\hat{x}}{\pa u}=0,\quad
t\frac{\pa\hat{u}}{\pa u}=\nu \hat{t}.
\label{eqc}\end{equation}
For $j=5$, $X_{5}=t\pa_{t}-\frac{2x}{h}\pa_{u}$
\begin{equation}
t\frac{\pa\hat{t}}{\pa t}-
\frac{2x}{h}\frac{\pa\hat{t}}{\pa u}=\hat{t},\quad
t\frac{\pa\hat{x}}{\pa t}-
\frac{2x}{h}\frac{\pa\hat{x}}{\pa u}=\sigma,\quad
t\frac{\pa\hat{u}}{\pa t}-
\frac{2x}{h}\frac{\pa\hat{u}}{\pa u}=-\frac{2\hat{x}}{h}.
\label{eqci}\end{equation}

The solution of equations (\ref{equ}), (\ref{eqd}),
and (\ref{eqt}) is;
\begin{equation}
\hat{t}=\mu t+\alpha,\quad \hat{x}=\mu\nu x+\beta,
\quad \hat{u}=\mu\nu u+\gamma. 
\end{equation}
Substituting in (\ref{eqc}) we get $\alpha=0$ and in (\ref{eqci})
we get $\sigma=0$ and $\beta =0$. The translation in $u$ is a
continuous symmetry so the possible discrete symmetries are:
\begin{equation}
\hat{t}=c_{1} t,\quad \hat{x}=c_{2} x, \quad \hat{u}=c_{2} u
,\quad c_{1},c_{2}\neq 0.
\end{equation}

Finally, we check if this transformation is a symmetry of the
equation.  To do so, we apply it to the Toda equation (\ref{toda}). 
The Toda equation in the new variables reads:
\begin{equation}\label{todanew}
    \frac{c_{1}^2}{c_{2}}\hat{u}_{\hat{t}\hat{t}}=
    \re^{\frac{1}{c_2}\left[\hat{u}\left(\hat{t},\hat{x}+
    \frac{\hat{h}}{c_{2}}\right)-\hat{u}(\hat{t},\hat{x})\right]}-
    \re^{\frac{1}{c_2}\left[\hat{u}\left(\hat{t},\hat{x}\right)-
    \hat{u}(\hat{t},\hat{x}-
    \frac{\hat{h}}{c_{2}})\right]}
    \end{equation}
with $\hat{h}=c_{2}h$. 
Then, $c_{2}=\pm 1,c_{1}=\pm 1$ are the only admissible solutions. 
When $c_{2}=1,c_{1}=-1$, we obtain a discrete symmetry, given by the
transformation:
\begin{equation}
\hat{t}=-t,\quad \hat{x}=x,
\quad \hat{u}=u.
\end{equation}
When $c_{2}=-1,c_{1}=1$, we get another discrete symmetry:
\begin{equation}
\hat{t}=t,\quad \hat{x}=-x,
\quad \hat{u}=-u.
\end{equation}

It is easy to check that the automorphism of Case b) 
cannot be realized in the representation under consideration.

%%%%%%%%%%%%%%%%%%%%%%%%%%%%%%%%%%%%%%%%%%%%%%%%%%%%%%%%%%%%%
%%%%%%%%%%%%%%%%%%%%%%%%%%%%%%%%%%%%%%%%%%%%%%%%%%%%%%%%%%%%%

\section{Conclusions}
In this article we have shown that the two methods, the automorphisms 
of the symmetry algebra \cite{Hy00a} and the determining equation for discrete 
symmetries \cite{GR96}, provide discrete symmetries, even in the case 
of discrete equations. It is worthwhile to notice that in the case of 
the Toda lattice, we have obtained  a non obvious discrete symmetry 
using these techniques. 

There are some drawbacks for both methods. In the case of the automorphism 
method, if the symmetry group is very large, the matrices involved 
become big and the final defining equations are very overdetermined 
and, in some cases may need symbolic manipulation programs to carry out the 
calculations. Moreover, the method is not applicable if there are no continuous 
symmetries, as it is the case, for example, of the Painlev\'e equations. 

In the case of the determining equation method, the equation for the 
function $F$ can some times be undetermined. For example, in the case of 
the Volterra equation
\begin{equation}\label{volt}
u_t=u(u_{+}-u_{-}),
\end{equation}
having chosen $F=F(t,u,u_{+},u_{-};\lambda)=u_{t}$ the determining
equation to solve is
\begin{equation}\label{voltdet}
\frac{\partial F}{\partial \lambda}+
\tau(t)\frac{\partial F}{\partial t}+
\phi(t,u)\frac{\partial F}{\partial u}+
\phi(t,u_{+})\frac{\partial F}{\partial u_{+}}+
\phi(t,u_{-})\frac{\partial F}{\partial u_{-}}=
\phi_{t}+(\phi_{u}-\tau')F.
\end{equation}    
We have no hint of the form of $\phi(t,u)$ and, consequently,  
equation (\ref{voltdet}) is not solvable. Of course, one could try to 
introduce some ansatzs for this function and maybe find some discrete 
symmetries. However, as one can easily show using the automorphism 
method, Volterra equation has no discrete symmetries and an ansatz 
will not provide any conclusion. 

Work is in progress to apply the other techniques mentioned at the 
beginning of Section 3 and for the construction of lattices and discrete
equations with prescribed discrete symmetries, a problem of interest
in Chemistry and Quantum Mechanics. Another topic which attracts our 
attention is the study of solutions of discrete equations invariant 
under a discrete symmetry group.

%%%%%%%%%%%%%%%%%%%%%%%%%%%%%%%%%%%%%%%%%%%%%%%%%%%%%%%%%%%%%
%%%%%%%%%%%%%%%%%%%%%%%%%%%%%%%%%%%%%%%%%%%%%%%%%%%%%%%%%%%%%

\section*{Acknowledgements}
This work was done while D L was visiting the Departamento de
F\'{\i}sica Te\'orica of Universidad Complutense de Madrid (Spain). 
His visit was financed by the {\it Ministerio of Educaci\'on, Cultura y
Deportes} of Spain (SAB01-0140).  This work was also partially
supported by Ministry of Science and Technology (Spain) under grant
BFM2002-02646 and NATO under grant PST.CLG.978431.

\end{document}